{
\documentclass{article}
 \relax
\setlength{\oddsidemargin}{.25in}
\setlength{\textwidth}{5.25in}
\setlength{\topmargin}{.25in}
\setlength{\textheight}{7.5in}
\usepackage{graphics}

\begin{document}

\title{WORMHOLE CORE, EXTRA DIMENSIONS, AND PHYSICAL UNIVERSE\footnote{This work is dedicated to one of my very bright students, Professor Naseem K. Rahman, on the occasion of his 60th birthday celebration. I am very proud of all his achievements and wish him every success in the future.} 
}         
\author{A. L. Choudhury   \\
Department of Chemistry and Physics  \\ Elizabeth City State University  \\
Elizabeth City, NC 27909 \\ email:alchoudhury@mail.ecsu.edu }      
\date{October 7, 2003}          
\maketitle

\begin{abstract}
   { We created a model of a several dimensional physical universe. The extra dimensions associated with the four dimensional physical universe is assumed to have a modified Gidding-Strominger wormhole core. This core is separated by a flexible wall, but it allows  the adiabatic pressure generated in the wormhole to penetrate in the extra dimensions. We assume that the extra dimensions are a contracting Robinson-Walker space. We show that the associated physical universe accelerates under a certain restriction of the parameters introduced. The extra dimensional space is  very large at the begining, however at present time this space will be very reduced. As a result the physical universe will appear to us four dimensional the way we observe it now.}
\end{abstract}

\section{Introduction}
   {$\;$$\;$In a recent paper [1] we have shown that we can reproduce an accelerating universe without introducing dark energy. We use a special case of a model introduced by Gu and Huang [2]. Instead of using their (3+n+1)-dimensional Robertson-Walker model, we took a seven dimensional space. Inside the extra 4-dimensional space-time which has a common time component with the real physical universe we live in, we introduce a modified Gidding-Strominger wormhole [3,4]. Although the wormhole solution has been obtained in Euclidean space-time we go over to the Lorentz space-time by analytically extending 't' from Euclidean space to 'it' in Lorentz space-time. Assuming such transition is possible, we obtain a time dependent scale factor. We than assumed that particles associated with the scalar field in the wormhole are expanding as an adiabatic gas and hence obtained the pressure in the wormhole as a function of time. Separating the wormhole from the extra dimensions by a flexible wall, we conjecture [5,6] that the pressure is transmitted into the extra dimensional space through the wall. We also improved the approximation for the time dependent pressure. This pressure we then identify as the pressure of the extra dimensional space. We then assume that the whole space has a Robertson-Walker space format. 
We derived  the the Einstein equations .
To solve those equations in the earlier paper we assumed that the extra dimensional space is expanding. In this paper,by contrast, we postulate that the extra dimensional  space is contracting. Although the unphysical wormhole started as a small blob inside the extra dimensions, the expanding wormhole swallows the contracting extra dimensions after a critical time $t_c$, making extra dimensions unvisible to the physical universe we live in. However, the effect of the  pressure still influences the structure of our physical four dimensional world. We will show in this paper that such physical universe expands and accelerates under certain condition of the parameters. 
$\;$$\;$In section 1, we show the basis of the wormhole model and derive the appropriate time dependence of the scale function a(t). In section 2,  we show how we obtain the pressure by assuming adiabatic expansion of the wormhole. We then develop the basis of the seven dimensional Robertson-Walker space in section 3. The coupled equations of the scale factors of the real physical world and the extra dimensional world have been derived there. In section 4 we conjecture that the extra dimension is contracting in a specific way. We then obtain the Hubble parameter for the physical universe.
$\;$$\;$ We compute the deacceleration parameter in section 5. We specify the condition under which the deacceleration parameter stays negative. We discuss our results in section 6.} 

\section{The unphysical inner wormhole core}

{$\;$$\;$The wormhole core which generates pressure on the Gu and Huang [2] universe with extra dimension is assumed to be obtained from the modified Gidding-Srominger model [3,4] as summerized in the papers by Choudhury and Pendharkar[5] , and Choudhury [6]. For the unphysical inner core we construct the model in an Euclidean space with time specified by $t_w$.  The core action in the Euclidean space is given as follows:    
\begin {equation}
S_E=\int {{d^4}x {{L_G}^c}(x)}+\int {{d^4}x {{L_{SA}}^c}(x)}=S_G+S_A,
\end{equation}
{where}
\begin {equation}
{{L_G}^c}(x)=\frac{{\surd g^c}{R(g^c)}} {2{\kappa}^2},
\end{equation}
{and}
\begin{equation}
{{L_{SA}}^c} (x)={\surd {g^c}}[{\frac1 2}(\bigtriangledown \Phi)^2+{g_p}^2 {\Phi}^2 Exp(\beta {\Phi}^2)]{H_{\mu\nu\rho}}^2.
\end{equation}
{In Eqs.(1) through (3)the suffix c stands for the core. The axion field is given by the relation }
\begin{equation}
H_{\mu\nu\rho}=\frac n {{g_p}^2 {a^3 (t)}} \varepsilon_{\mu\nu\rho},
\end{equation}
{ from which we get}
\begin{equation}
H^2=\frac{6D} {{g_p}^2 {a^6}},
\end{equation}
{with}
\begin{equation}
D=\frac{n^2} {{g_p}^2}
\end{equation}
{The space-time interval in Euclidean space is given by}
\begin {equation}
ds^2=d{t_w}^2+a^2(t)(d\chi^2+sin^2 \chi d\theta^2+sin^2 \chi sin^2 \phi d\phi^2).
\end{equation}
{The variation of $g_{\mu\nu}^c$ in the core leads to the following equation}
\begin{equation}
R_{\mu\nu}^c-{\frac1 2}g_{\mu\nu}^c R^c=\kappa^2[\nabla_\mu \Phi \nabla_\nu \Phi-({\frac1 2}(\nabla\Phi)^2) g_{\mu\nu}^c+{g_p}^2 Exp(\beta {\Phi}^2)({H_\mu}^\alpha\gamma H_\nu\alpha\gamma- {\frac1 6} g_{\mu\nu}^c H_{\alpha\beta\gamma} ^2)]
\end{equation}
{For the $\Phi$-variation inside the core the equation of motion of $\Phi$ yields}
\begin{equation}
{\nabla}^2 \Phi-2{{g_p}^2\beta}\Phi{Exp(\beta\Phi^2)}{H_{\alpha\beta\rho}}^2=0.
\end{equation}
{ The Hamiltonian constraint yields}
\begin{equation}
(\frac1 a \frac{da} {dt_w})^2-\frac1 {a^2}=\frac{\kappa^2} 3 [{\frac1 2}(\frac{d\Phi} {dt})^2-6Exp(\beta\Phi^2) \frac{n^2} {g_p^2 a^6}].
\end{equation}
{The dynamical equation yields }
\begin{equation}
{\frac d {dt_w}} ({\frac1 a }{\frac{da} {dt_w}})+\frac1 {a^2}=-\kappa^2[(\frac{d\Phi} {dt_w})^2-12 Exp(\beta\Phi^2) {\frac{n^2} {g_p^2 a^6}}].
\end{equation}
{ Now we substitute  a new variable $\tau$ defined by the relation}
\begin{equation}
d\tau=a^{-3} dt_w.
\end{equation}
{Combining these equations we can derive the equation satisfied by the scale function for the wormhole core as }

\begin{equation}
{(\frac1 a }{\frac{da} {dt_w}})^2-a^4+{a_c}^4=0,
\end{equation}
{where}
\begin{equation}
a_c=\surd(\frac{\kappa^2C_o} 2)
\end{equation}
{and $C_o$ is a constant. }
{$\;$$\;$The solution of the Eq.(13) has been obtained by Gidding and Strominger and is given by}
\begin{equation}
a^2(\tau)={a_c}^2\surd(sec(2{a_c}^2\tau)).
\end{equation}
{The scale factor can now be of two possible forms }
\begin{equation}
a(\tau)=\pm a_c (sec(2{a_c}^2\tau))^{\frac1 4},
\end{equation}
{specified by the signs. Both the solutions stand on equal footing.}
\section{Pressure generated by the wormhole}
{$\;$$\;$ Following suggestions originated and further developed by Choudhury and Pendharkar [2] we assume that the wormhole is in a gaseous state satisfying the adiabatic gas law }   
\begin{equation}
P_w V_w ^\gamma=Constant=B_1,
\end{equation}
{where $\gamma$  is a constant. The volume of the wormhole can be shown to be proportional to $a^3 (\tau)$ . The pressure can be expressed as }
\begin{equation}
P_w={(\pm 1)^{-3\gamma}}B{a_c}^{-3\gamma}[sec(2{a_c}^2 \tau)]^{-(3\gamma/4)}
\end{equation}
{where B is a new constant. We here, in contrast to previous manipulation, assume that $\tau$ can be identified with real time t of the physical world. We can thus write }

\begin{equation}
P_w={(\pm 1)}^{-3\gamma}B{a_c}^{-3\gamma}[sec(2a_c^2t)]^{(-3\gamma/4)}.
\end{equation}
{Assuming a special value of $\gamma$ for the unphysical wormhole to to be $\gamma=8/3$}
\begin{equation}
P_w=B{a_c}^{-8}cos^2(2a_c^2t).
\end{equation}
\section{ Seven dimensional universe and Hubble parameter}
{$\;$$\;$We here introduce a seven dimensional physical universe, a special case of the Gu and Huang model [2]. This space we assume to be a homogeneos and isotropic Robertson-Walker space. The interval is given by the relation}  

\begin{equation}
d {s_P}^2 = -d t^2 + d {s_\alpha}^2+ d {s_\beta}^2
\end{equation}
{where }
\begin{equation}
d {s_\alpha}^2={{\alpha^2}(t)}[\frac{d{r_\alpha}^2} {1-{k_\alpha}{r_\alpha}^2} + {{r_\alpha}^2}d {\theta_\alpha}^2 + {{r_\alpha}^2}{sin^2}{\theta_\alpha} d{\phi_\alpha}^2],
\end{equation}
{and}
\begin{equation}
d {s_\beta}^2={{\beta^2}(t)}[\frac{d{r_\beta}^2} {1-{k_\beta}{r_\beta}^2} + {{r_\beta}^2}d {\theta_\beta}^2 + {{r_\beta}^2}{sin^2}{\theta_\beta} d{\phi_\beta}^2]
\end{equation}

{we have chosen c=1. The Einstein equation is as follows}

\begin{equation}
G_{\mu\nu}=R_{\mu\nu}-{\frac1 2}g_{\mu\nu} R  = - {8 \pi G}{T_{\mu\nu}}.
\end{equation}

{In the above equation $\mu$ and $\nu$ run from 0 through 6. The tensor $T_{\mu\nu}$ can be defined as}
\begin{equation}
T_{\mu\nu}= G {T_{\mu\nu}}^{(\alpha)}+ G' {T_{\mu\nu}}^{(\beta)},
\end{equation}
{where}
\begin{equation}
{T_{\mu\nu}}^{(i)}= p_i {g_{\mu\nu}}^{(i)}+(p_i +\rho_i) {U_\mu}^i {U_\nu}^i,
\end{equation}
{with $i=\alpha $and $\beta$.
In the above expressions}
\begin{equation}
{g_{\mu\nu}}^{(\alpha)}=-1,  for  \mu=\nu=0;
\end{equation}
\begin{equation}
{g_{\mu\nu}}^{(\alpha)}=1,  for  \mu=\nu=1,2,3;
\end{equation}
\begin{equation}
{g_{\mu\nu}}^{(\alpha)}=0,  for  \mu\not=\nu=1,2,3;
\end{equation}
{and}
\begin{equation}
{g_{\mu\nu}}^{(\alpha)}=0,  for  \mu=\nu=4,5,6.
\end{equation}
{$\;$$\;$Similarly}
\begin{equation}
{g_{\mu\nu}}^{(\beta)}=-1,  for  \mu=\nu=0;
\end{equation}
\begin{equation}
{g_{\mu\nu}}^{(\beta)}=-1,  for  \mu=\nu=4,5,6;
\end{equation}
\begin{equation}
{g_{\mu\nu}}^{(\beta)}=0,  for  \mu\not=\nu=1,2,3;
\end{equation}
{and}
\begin{equation}
{g_{\mu\nu}}^{(\beta)}=0,  for  \mu=\nu=1,2,3.
\end{equation}
{$\;$$\;$The equation}
\begin{equation}
G_{tt}=-8\pi(G{T_{tt}}^{(a)}+G'{T_{tt}}^{(b)}
\end{equation}
{turns into the form}
\begin{equation}
3[\frac{(\dot\alpha(t))^2+k_\alpha } {(\alpha(t))^2}]+\frac{(\dot\beta(t))^2+k_\beta } {(\beta(t))^2}+2{\frac{\dot\alpha(t)} {\alpha(t)}} {\frac{\dot\beta(t)} {\beta(t)}}=-8\pi(G{\rho_\alpha}+G'{\rho_\beta})=-8\pi\rho.
\end{equation} 
{ Similarly we get for the  ${r_a}$-${r_a}$ component}                                                                                                                                                                                                                                                                            
\begin{equation}
2\frac{\ddot{\alpha}(t)}  {\alpha(t)}+3\frac{\ddot{\beta}(t)} {\beta(t)}+3\frac{(\dot\alpha(t))^2+k_\alpha } {(\alpha(t))^2}+3\frac{(\dot\beta(t))^2+k_\beta } {(\beta(t))^2}+6{\frac{\dot\alpha(t)} {\alpha(t)}} {\frac{\dot\beta(t)} {\beta(t)}}= 8{\pi} G{p_\alpha}.
\end{equation}
{Both $\theta_\alpha$-$\theta_\alpha$ and $\phi_\alpha$-$\phi_\alpha$ component equations are given by}
\begin{equation}
2\frac{\ddot{\alpha}(t)}  {\alpha(t)}+3\frac{\ddot{\beta}(t)} {\beta(t)}+\frac{(\dot\alpha(t))^2+k_\alpha } {(\alpha(t))^2}+3\frac{(\dot\beta(t))^2+k_\beta } {(\beta(t))^2}+3{\frac{\dot\alpha(t)} {\alpha(t)}} {\frac{\dot\beta(t)} {\beta(t)}}= 8{\pi}G{p_\alpha}.
\end{equation}
{$\;$$\;$For the ${r_\beta}$-${r_\beta}$ component we get}
\begin{equation}
3\frac{\ddot{\alpha}(t)}  {\alpha(t)}+2\frac{\ddot{\beta}(t)} {\beta(t)}+3\frac{(\dot\alpha(t))^2+k_\alpha } {(\alpha(t))^2}+3\frac{(\dot\beta(t))^2+k_\beta } {(\beta(t))^2}+6{\frac{\dot\alpha(t)} {\alpha(t)}} {\frac{\dot\beta(t)} {\beta(t)}}= 8{\pi} G'{p_\beta}.
\end{equation}
{For $\theta_b$-$\theta_\beta$ and $\phi_\beta$-$\phi_\beta$ components we obtain}
\begin{equation}
3\frac{\ddot{\alpha}(t)}  {\alpha(t)}+2\frac{\ddot{\beta}(t)} {\beta(t)}+3\frac{(\dot\alpha(t))^2+k_\alpha } {(\alpha(t))^2}+\frac{(\dot\beta(t))^2+k_\beta } {(\beta(t))^2}+3{\frac{\dot\alpha(t)} {\alpha(t)}} {\frac{\dot\beta(t)} {\beta(t)}}= 8{\pi} G'{p_\beta}.
\end{equation}
{ In this paper we set $k_\alpha$=$k_\beta$=0. The above equations then changes into the form}
\begin{equation}
\frac{(\dot\alpha(t))^2 } {(\alpha(t))^2}+\frac{(\dot\beta(t))^2 } {(\beta(t))^2}+2{\frac{\dot\alpha(t)} {\alpha(t)}} {\frac{\dot\beta(t)} {\beta(t)}}=-{\frac8 3}\pi\rho,
\end{equation} 
\begin{equation}
2\frac{\ddot{\alpha}(t)}  {\alpha(t)}+3\frac{\ddot{\beta}(t)} {\beta(t)}+3\frac{(\dot\alpha(t))^2 } {(\alpha(t))^2}+3\frac{(\dot\beta(t))^2 } {(\beta(t))^2}+6{\frac{\dot\alpha(t)} {\alpha(t)}} {\frac{\dot\beta(t)} {\beta(t)}}= 8{\pi} G{p_\alpha},
\end{equation}
\begin{equation}
3\frac{\ddot{\alpha}(t)}  {\alpha(t)}+2\frac{\ddot{\beta}(t)} {\beta(t)}+3\frac{(\dot\alpha(t))^2 } {(\alpha(t))^2}+\frac{(\dot\beta(t))^2 } {(\beta(t))^2}+3{\frac{\dot\alpha(t)} {\alpha(t)}} {\frac{\dot\beta(t)} {\beta(t)}}= 8{\pi} G'{p_\beta},
\end{equation}
\begin{equation}
2\frac{\ddot{\alpha}(t)}  {\alpha(t)}+3\frac{\ddot{\beta}(t)} {\beta(t)}+\frac{(\dot\alpha(t))^2 } {(\alpha(t))^2}+3\frac{(\dot\beta(t))^2 } {(\beta(t))^2}+3{\frac{\dot\alpha(t)} {\alpha(t)}} {\frac{\dot\beta(t)} {\beta(t)}}= 8{\pi}G{p_\alpha},
\end{equation}
{and}
\begin{equation}
3\frac{\ddot{\alpha}(t)}  {\alpha(t)}+2\frac{\ddot{\beta}(t)} {\beta(t)}+3\frac{(\dot\alpha(t))^2} {(\alpha(t))^2}+\frac{(\dot\beta(t))^2} {(\beta(t))^2}+3{\frac{\dot\alpha(t)} {\alpha(t)}} {\frac{\dot\beta(t)} {\beta(t)}}= 8{\pi} G'{p_\beta}.
\end{equation}
\section{Hubble parameter}
{$\;$$\;$In this section we intend to derive the Hubble parammeter of our 4-dimensional physical universe associated with the scale factor $\alpha(t)$. At this step we would assign certain properties to the extra dimension which are in contrast to the assumptions we made in a previous paper. We would assume here that the extra dimensions are contracting whereas the unphysical wormhole is expanding.
$\;$$\;$Subtracting from Eq.(45) the Eq.(43) we get}
\begin{equation}
\frac{(\dot\beta(t))^2 } {(\beta(t))^2}+3{\frac{\dot\alpha(t)} {\alpha(t)}}\frac{(\dot\beta(t))} {(\beta(t))}=0.
\end{equation} 
{Combining Eqs.(46) and (41) we get}
\begin{equation}
\frac{(\dot\alpha(t))^2 } {(\alpha(t))^2}+{\frac1 2}{\frac{\dot\alpha(t)} {\alpha(t)}}{\frac{\dot\beta(t)} {\beta(t)}}=-8{\pi}{\rho}.
\end{equation} 
{Since the  extra dimensions are introduced to adjust the real universe with co-ordinates t, $r_\alpha$,$\theta_\alpha$,and $\phi_\alpha$, we make the conjecture here in contrast to the previous assumption that the space with extra dimensions with coordinates, t,$r_\beta$,$\theta_\beta$,and $\phi_\beta$  contracts at a constant rate. That is,we assume that $\beta(t)$  satisfies the relation}
\begin{equation}
\frac{(\dot\beta(t)) } {(\beta(t))}= -4\Delta=H_\beta(t).
\end{equation}
{where $\Delta$ is a constant. The special case, which $\beta(t)$ satisfies, can be written as}
\begin{equation}
\beta(t)={\beta(0)}{e^{-4\Delta t}}
\end{equation}
{Using this relation we can write the Eq.(47) into the following form:}
\begin{equation}
\frac{(\dot\alpha(t))^2 } {(\alpha(t))^2}-2\Delta{\frac{\dot\alpha(t)} {\alpha(t)}}+8{\pi}{\rho}=0.
\end{equation}
{Writing } 
\begin{equation}
H_{\alpha}=\frac{\dot\alpha(t)} {\alpha(t)},
\end{equation}
{we get from Eq.(50)}
\begin{equation}
H_{\alpha}=\Delta\pm\surd({\Delta}^2+8\pi\rho).
\end{equation}
{$\;$$\;$Since $H_\alpha$ is the Hubble parameter of the real world we live in with coordinates t,$r_a$,$\theta_a$,and $\phi_a$ to match the expanding universe, we discard the solution with the negative sign before the square root. Therefore we get for ${H_\alpha}(t)$}
\begin{equation}
H_{\alpha}=\Delta+\surd({\Delta}^2+8\pi\rho).
\end{equation}
{Since G, G', $\rho_\alpha$, $\rho_\beta$ are assumed to be positive we find that}
\begin{equation}
H_{\alpha}>0.
\end{equation}
{The time dependence of the Hubble parameter lies in the fact that the densities of the real and extra dimensions may depend on time.}
\section{The deaccelerating parameter of the real world}
{$\;$$\;$From the Eqs(42) and (43) we get}
\begin{equation}
\frac{\ddot{\alpha}(t)}  {\alpha(t)}-\frac{\ddot{\beta}(t)} {\beta(t)}=8\pi(G'p_\beta-Gp_\alpha)
\end{equation}
{On account of the Eq.(48) we find}
\begin{equation}
\frac{\ddot{\alpha}(t)}  {\alpha(t)}= 16{\Delta}^2 + 8\pi(G'p_\beta-Gp_\alpha)
\end{equation}
{Looking back on our construction of the universe, we assumed that the wormhole is placed at the center of the extra dimensions. The extra dimensional space is contracting  at the rate of $H_b(t)$. The wormhole pressure is transferred through the wall of the wormhole to the extra dimensions. If we now assume that there is no extra pressure except the pressure transferred through the wall to the extra dimension,then $p_b=P_w$. We get then the deacceleration parameter $q_0$ to be}
\begin{equation}
{q_0}(t_0)=-{\frac{\ddot{\alpha}(t)}  {\alpha(t)}}{\frac1 {{H^2}(t_0)}}
=-[16{\Delta}^2 + 8\pi(G'P_w-Gp_\alpha)]{\frac1 {{H^2}(t_0)}},
\end{equation} 
{where $t_0$ represents present time (see Choudhury and Pendharkar [5]).

$\;$$\;$With our special choice $\gamma=\frac8 3$, we can use the expression for $P_w$ of the Eq.(20). Therefore we get}
\begin{equation}
q_0(t_0)=-[16{\Delta}^2+8\pi[G'B{{a_c}^{-8}}cos^2(2{a_c}^2t)-Gp_a]].
\end{equation}
{If the expansion rate $\Delta$ is assumed to satisfy a relation}
\begin{equation}
16{\Delta}^2-8\pi G p_a \geq 8\pi G' B{{a_c}^{-8}}.
\end{equation}
{and we get}
\begin{equation}
q_0(t_0)\leq -8\pi G'B{{a_c}^{-8}}[1+cos^2(2{a_c}^2t) ].
\end{equation}
{The above quantity $q_0$ is always negative. Therefore the universe is accelerating [7,8]. If we take an extreme case where the equality sign in Eq.(6.6) holds, we get for $q_0$ }
\begin{equation}
q_0(t_0)=-8\pi G'B{{a_c}^{-8}}[1+cos^2(2{a_c}^2t)].
\end{equation}
{This leads to a spectacular outcome. The deacceleration parameter fluctuates with time. It increases and decreases periodically. If such fluctuation is detected in future observation, our model will be a viable one.} 

\section{Concluding remarks}
{$\;$$\;$Following Gu and Huang, we have constructed a model introducing extra dimensions. Our one is a special case of that model where we have incorporated only three extra dimension.  However, we have incorporated an expanding modified Gidding-Strominger wormhole at the center of the extra dimensions. This wormhole generates an adiabatic pressure. A flexible wall separates the wormhole from the extra dimensional space. This pressure influences the deacceleration parameter of our expanding universe. Introducing certain restriction on the parameters we have shown that the observational outcome of the accelerating universe can be reproduced. 
$\;$$\;$However this model has some extra suitable characteristics. We start from a seven dimensional model where in the beginning all seven dimensions were observable. We incorporated a  modified Gidding-Strominger wormhole at the center of the extra dimensions. Since the extra dimensions keep on contracting at a critical time the unphysical wormhole swallows the extra dimensions making it invisible. Beyond a critical time we thus only see the four dimensions. However, under special    restrictions we find that the physical dimensions accelerate with regular fluctuations. If future observations show such fluctuations then our model will be validated. }

\section{References}
\begin{enumerate}
\item{A. L. Choudhury: Influence on the physical universeby wormhole generated extra dimensional space; arXiv.gr-qc/0311043v1,13 Nov 2003.}
\item {Je-An Gu and W-Y. P. Huang, arXiv:astro-ph/0112565 v1 31 Dec 2001.}
\item { S. B. Giddings and A.  Strominger , Nucl. Phys. B 307, 854 (1988).}
\item { D. H. Coule and K. Maeda, Class. Quant. Grav. 7, 955 (1990).}
\item { A. L. Choudhury, Hadronic J., 23, 581 (2000).}
\item { L. Choudhury and H. Pendharkar, Hadronic J. 24, 275 (2001).}

\item { N. Bahcall, J. P. Ostriker, S. Perlmutter, and P. J. Steinhardt, Science, 284,1481 (1999).}
\item { C. Aremendariz Picon, V. Mukhanov and Paul J. Steinhardt: Essentials of k-Essence. ArXiv:astro-ph/0006373 (2000).}
\end{enumerate}    


\end{document}